\documentclass[10pt]{bmc_article}

\usepackage{cite} 
\usepackage{url}  
\usepackage{ifthen}  
\usepackage{multicol}   
\usepackage[utf8]{inputenc} 
\usepackage{setspace}
\usepackage{epsfig}
\usepackage{graphicx}
\usepackage{amsmath, amsthm, amssymb}
\usepackage{subfig}
\usepackage{caption}
\usepackage{multirow}
\usepackage{algorithmic}

\newtheorem{theorem}{Theorem}
\newtheorem{lemma}{Lemma}

\newtheorem{corollary}{Corollary}

\newboolean{publ}

\begin{document}

\begin{singlespace}

\title{Improved Algorithms for Approximate String Matching (Extended Abstract)}

\author{Dimitris Papamichail\correspondingauthor$^1$%
       \email{Dimitris Papamichail\correspondingauthor - dimitris@cs.miami.edu}%
       and
         Georgios Papamichail$^2$%
         \email{Georgios Papamichail - pmichael@ekdd.gr}%
}%


\address{%
    \iid(1)Department of Computer Science, University of Miami, Coral Gables, Miami, USA\\
    \iid(2)National Center of Public Administration, Athens, Greece%
}%

\maketitle

\begin{multicols}{2}

\section*{Abstract}
        \paragraph*{Background:} 
The problem of approximate string matching is important in many different areas such
as computational biology, text processing and pattern recognition. A great effort has
been made to design efficient algorithms addressing several variants of the problem, 
including comparison of
two strings, approximate pattern identification in a string or
calculation of the longest common subsequence that two strings share.
\\
        \paragraph*{Results:} 
We designed an output sensitive algorithm solving the edit distance problem between two strings of
lengths $n$ and $m$ respectively in
time $O( (s-|n-m|)min(m,n,s)+m+n)$ and linear space, where $s$ is the edit distance
between the two strings. This worst-case time bound sets the quadratic factor of the algorithm
independent of the longest string length and improves existing theoretical bounds for this problem.
The implementation of our algorithm excels also in practice, especially in cases
where the two strings compared differ significantly in length.
\\
        \paragraph*{Conclusions:} 
We have provided the design, analysis and implementation of a new algorithm for calculating
the edit distance of two strings with both theoretical and practical implications. Source code
of our algorithm is available at http://www.cs.miami.edu/\~{ }dimitris/edit\_distance.

\section*{Background}

Approximate string matching is a fundamental, challenging problem in 
Computer Science, often requiring a large amount of computational 
resources. It finds applications in different areas such as computational 
biology, text processing, pattern recognition and signal processing. For 
these reasons, fast practical algorithms for approximate string matching 
are high in demand. There are several variants of the approximate string
matching problem, including 
the problem of finding a pattern in a text allowing a limited number of 
errors and the problem of finding the number of edit operations that can 
transform one string to another. We are interested in the latter form 
in this paper.

The edit distance $D(A,B)$ between two strings $A$ and $B$ is defined in
general as the minimum cost of any sequence of edit operations that edits
$A$ into $B$ or vice versa. In this work we will focus on the Levenshtein
edit distance \cite{lev65}, where the allowed edit operations are
insertion, deletion or substitution of a single character, with each 
operation carrying a cost of $1$. This type of operation is often called
unit-cost edit distance and is considered the most common form. The weighted
edit distance allows the same operations as the Levenshtein edit distance, but
each operation may have an arbitrary cost.

\begin{figure*}[htbp]
\centerline{
\epsfig{file=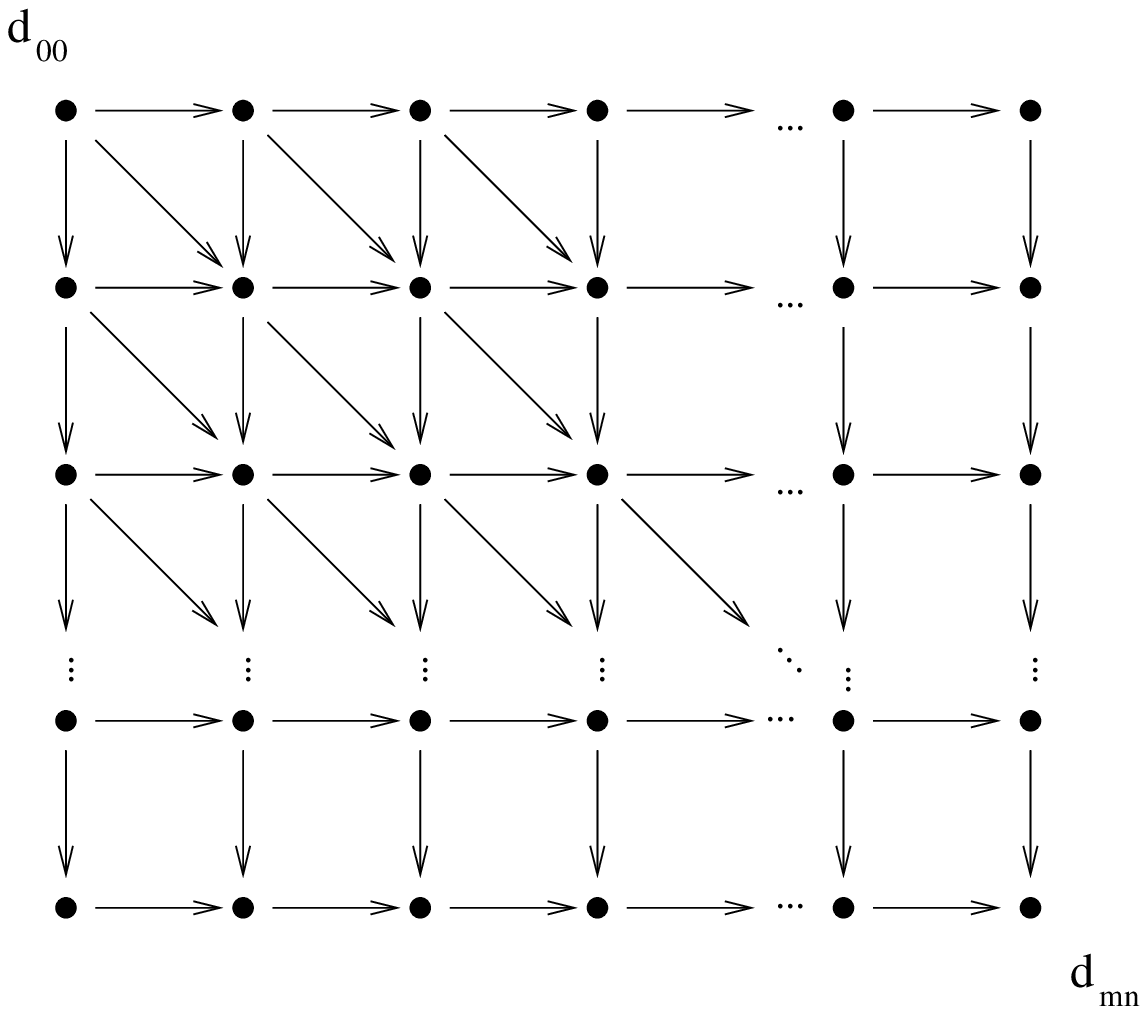, width=6cm}
}
\caption{Dependency graph}
\label{fig:dependency_graph}
\end{figure*}

In the literature there exist a number of algorithms dealing with the calculation of the edit distance
between two strings. The basic dynamic programming algorithm that solves
the problem in $O(mn)$ time and linear space has been invented and analyzed several
times in different contexts \cite{vin68,nee70,san72,sel74,wag74,low75}, published between 1968 and 1975. Early
on there was an algorithm by Masek and Paterson \cite{mas80}, building on a
technique called the "Four-Russian paradigm" \cite{arl70}, which computes
the edit distance of two strings over a finite alphabet in time $O(mn/\log^2n)$. This algorithm is not
applicable in practice, since it can outperform the basic algorithm only
then the input size is exceeding 40GB. All these algorithms can also be
used to calculate the {\it alignment} of two strings, in addition to their edit
distance. A modification of the basic algorithm by Hirschberg \cite{hir75} allows
the alignment calculation to be performed using linear space as well.

A few years later in 1985, Ukkonen arrived at an $O(s\cdot min(m,n))$ time algorithm,
using space $O(min(m,n,s))$ \cite{ukk85}, where $s$ is the edit distance of the two
strings compared, creating a very efficient output sensitive algorithm for this problem. 
The following year, Myers published an
algorithm for the {\it Longest Common Substring (LCS)} problem, which is
similar to the edit distance problem, which has $O(s^2 + (m+n)\log(m+n))$
time and linear space complexity \cite{mye86}. In achieving this result,
a generalized suffix tree of the input strings, supplemented by {\it Lowest Common Ancestor (LCA)} information,
has to be used, which renders the solution impractical and only of theoretical value.
The practical version of that algorithm needs $O(s(m+n))$ time. From the other hand, a variation of Ukkonen's 
algorithm using $O(s\cdot min(s,m,n))$ space leads to an efficient,
straightforward implementation, using recursion. Lastly, the basic algorithm,
although theoretically inferior, is the most commonly used,
owing to its adaptability, ease of implementation, instruction value,
and speed, the latter being a result of low constant operations.

In this paper we will present an $O((s-|n-m|)\cdot min(m,n,s) + m + n)$ time and linear space algorithm
to calculate the edit distance of two strings, which
improves on all previous results, the implementation of which is practical and competitive
to the fastest algorithms available. The quadratic factor in the time complexity
now becomes independent of the longest string, with the algorithm performing its best
when the two strings compared differ significantly in size.

\section*{Methods}

\subsection*{Definitions}

In this section we closely follow the notation and definitions
in Ref. \cite{ukk85}.
Let $A = a_1 a_2 \ldots a_n$ and $B = b_1 b_2 \ldots b_m$ be two
strings of lengths $n$ and $m$ respectively, over a finite alphabet
$\Sigma$. Without loss of generality, let $n \ge m$.

To define the edit distance between two strings $A$ and $B$, we will
let the possible editing operations be {\it deletion}, {\it insertion} and
{\it substitution}. Then we define {\it edit distance} as the minimum number of character
insertions, deletions and substitutions to transform string $A$ to string $B$. 
By that definition, each editing operation has a cost of $1$ and this
edit distance in bibliography is usually referred to as Levenshtein edit distance \cite{lev65}.
Edit operations can be generalized to have non-negative costs,
but for the sake of simplicity in the analysis of our algorithm we will
concern ourselves only with the Levenshtein edit distance.
We also assume that there is
always an editing sequence with cost $D(A,B)$ converting $A$ into $B$ such
that if an cell is deleted, inserted or changed, it is not modified again.
Under these assumptions the edit distance is symmetric and it holds $0 \le s \le max(n,m)$.
Since $n \ge m$ and there is a minimum number of $n-m$ indels that need to be applied
in transforming $A$ into $B$, the last equation becomes $n-m \le s \le n$.
The insertion and deletion operations are symmetric, since an insertion, when
transforming $A$ to $B$, is equivalent to a deletion in the opposite transformation,
and vice versa. Both operations will be referred to as {\it indels}.

The basic dynamic programming algorithm employed to solve the edit distance
problem, invented in a number of different contexts \cite{vin68,nee70,san72,sel74,wag74,low75}, 
makes use of the edit graph, an $(n+1)\times(m+1)$ matrix
$(d_{ij})$ that is computed from the recurrence:

\begin{center}
\small
\begin{tabular}{rcll}
$d_{00}$&$=$&$0$&\\
$d_{ij}$&$=$&min(&$d_{i-1,j-1} + ($If $a_i = b_j$ then $0$ else $1)$,\\
&&& $d_{i-1,j} + 1$,\\
&&& $d_{i,j-1} + 1)$, $i > 0$ or $j > 0$.
\end{tabular}
\end{center}

This matrix can be evaluated starting from $d_{00}$ and proceeding row-by-row or column-by-column.
This process takes time and space $O(mn)$ and produces the edit distance of the strings in position $d_{mn}$.
The cells of the matrix (nodes of the graph) have dependencies based on this recurrence, forming
the {\it dependency} or {\it edit graph}, a directed acyclic graph that is shown 
in Fig.\ref{fig:dependency_graph}. All edit graph nodes will be referred to
as {\it cells} and all graph edges (edit operations) will be referred to as {\it transitions}. 

To refer to the diagonals of $(d_{ij})$ we number them with integers $-m, -m+1,\ldots,0,1,\ldots,n$
such that the diagonal denoted by $k$ consists of those $d_{ij}$ cells for which $j-i=k$. The diagonal
$n-m$, where the final value $d_{mn}$ resides, is special for our purposes and we will call it 
{\it main diagonal}. The matrix cells between diagonals $0$ and $n-m$ (inclusive) consist the {\it center}
of the edit graph/matrix, the lower left triangle between diagonals $-1$ to $-m$ will be called the {\it left corner}
of the graph and upper right triangle between diagonals $n-m+1$ and $n$ will be called the
{\it right corner} of the graph.

A {\it path} in the edit graph is a series of transitions connecting cells, similar to a path in a directed
graph.
Whenever we generally refer to a path, we will assume that the final cell it reaches is $d_{mn}$. The {\it optimal path} will be
a path originating at $d_{00}$, and for which the sum of the costs of its transitions is minimal among all paths from $d_{00}$.

\subsection*{The concept}

\begin{figure*}[htbp]
\centerline{
\subfloat[Basic algorithm scoring scheme]{\epsfig{file=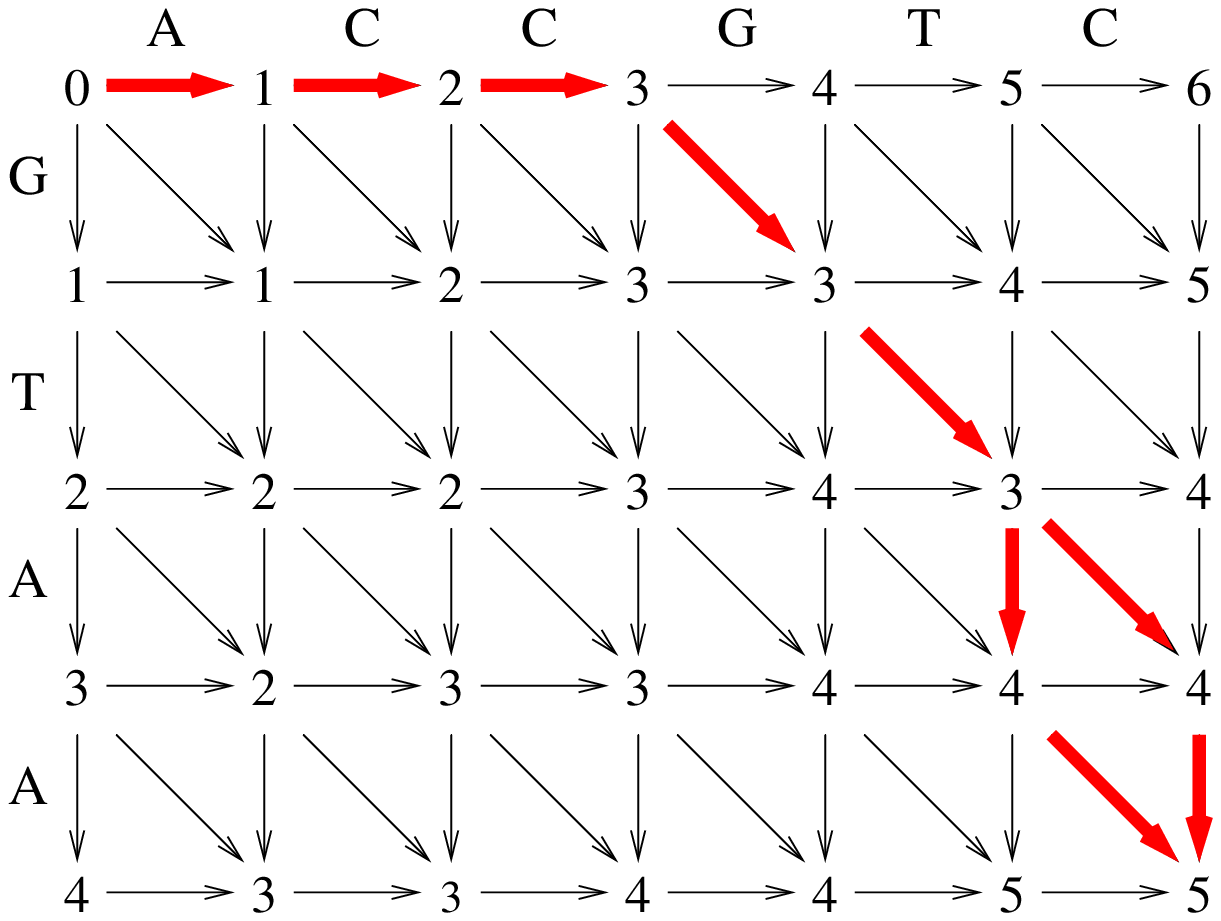, width=0.35\textwidth}}
\hspace{10mm}
\subfloat[New scoring scheme]{\epsfig{file=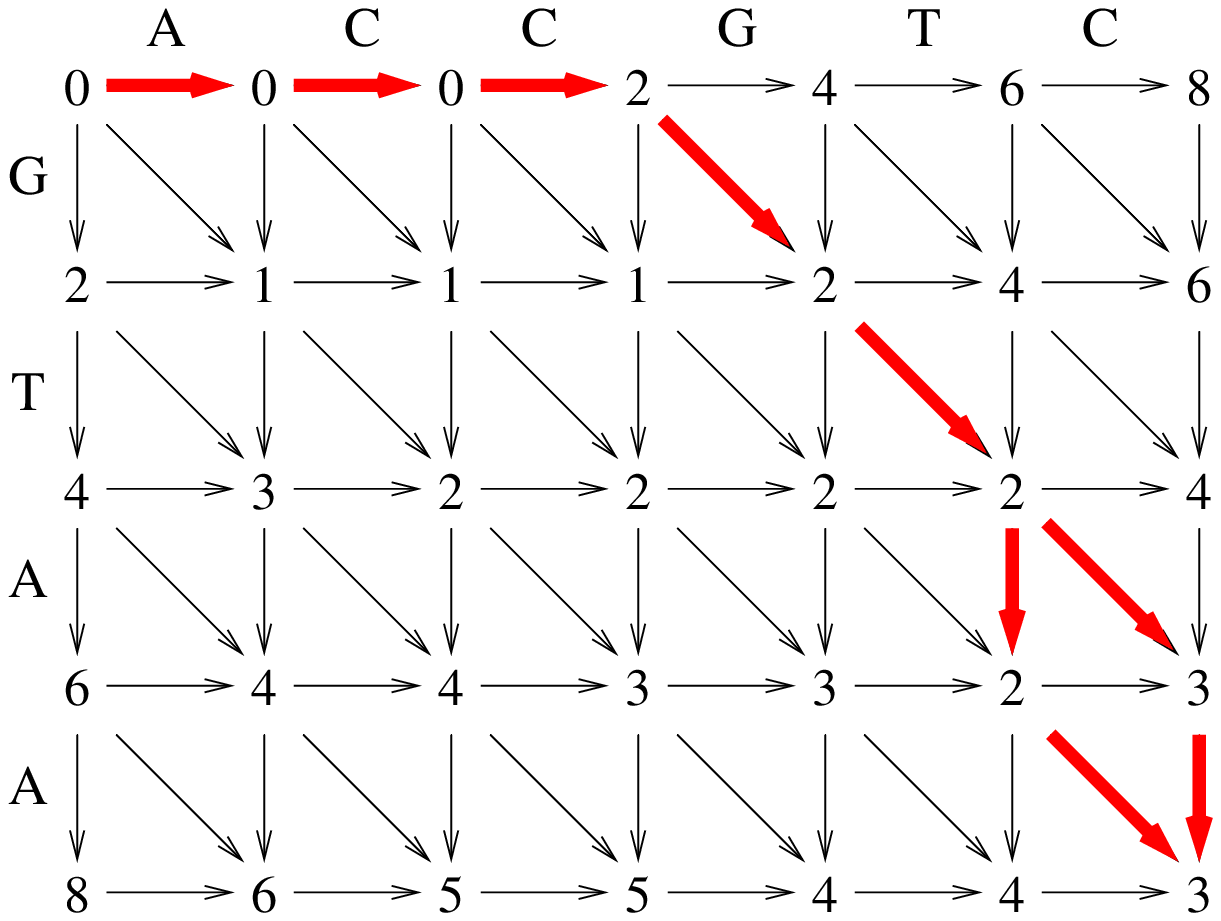, width=0.35\textwidth}}
}
\caption{Edit graph cell values and optimal paths under different scoring schemes}
\label{fig:score}
\end{figure*}

The basic dynamic programming algorithm evaluates unnecessary values $d_{ij}$. This observation led Ukkonen
\cite{ukk85} design an algorithm that is diagonal-based and computes cell values only between the diagonals
$-s$ and $n-m+s$. He also used the observation that, under the edit operations' cost scheme discussed previously,
the values in a diagonal are monotonically increasing, where for Levenshtein edit distance costs it
is furthermore $d_{i+1,j+1} \in \{d_{i,j},d_{i.j}+1\}$. 

Both Ukkonen \cite{ukk85}, for calculating the edit distance, and Myers \cite{mye86}, for calculating
the length of the {\it Longest Common Substring} (LCS) of two strings, a problem closely related to the edit distance, 
designed their
algorithms with a common feature: The iterations in evaluating the edit graph cells were score based,
as opposed to column or row based in the basic algorithm. In each step they would increase the edit 
distance $D$ by $1$, starting at $0$, and evaluate all cells with values $d_{ij} \le D$, meaning cells reachable with
edit distance $D$, often omitting cells not contributing to the next iteration, by considering
transitions between cells where the values are incremented.

The algorithm we present here builds on all previous observations and the main iteration is score based as
well. But we also make use of the following facts:

\begin{enumerate}
\item A number of indels ($n-m$) is unavoidable when the two strings considered differ in size.
\item Additional indels are unavoidable when the optimal path strays away from the main diagonal.
\item Certain cells do not contribute to the optimal path or their contribution is redundant.
\end{enumerate}

The first point is obvious, since the optimal path, starting at diagonal $0$ and ending at diagonal $n-m$,
can only use indels to progress through diagonals. To argue towards the second fact we notice that
every time a path follows an indel from diagonal $k$ to $k-1$ when $k \le n-m$ or from diagonal $k$ to $k+1$
when $k \ge n-m$, that path will need to follow another compensatory indel at some later point, in order to reach the 
main diagonal, where the target cell $d_{mn}$ resides.

In order to address the third fact, we will introduce the concept of {\it dominance}. We will say that cell
$d_{ij}$ dominates cell $d_{kl}$ if no path through $d_{kl}$ defines a better edit distance than the 
optimal path through $d_{ij}$. This implies that $d_{ij}$ has an equal or better potential to belong
to the optimal path (which defines $s$) than $d_{kl}$, and thus the latter and its paths do not need to
be considered further.

Some dominance relations between cells can be spotted easily. Let us consider all possible paths starting
from $d_{00}$. If a match exists between characters $a_1$ and $b_1$ ($a_1 = b_1$), then we 
do not need to consider indel transitions from $d_{00}$ to $d_{10}$ and $d_{01}$. In that case actually,
all cells $d_{0k}$ for $1 \le k \le n$ and $d_{k0}$ for $1 \le k \le m$ are dominated by $d_{11}$.
Since $a_1$ matches $b_1$, cell $d_{11}$ obtains the value of $0$. Then all cells $d_{1k}$,
$2 \le k \le n$ can obtain a value of $k-1$ through a path traversing $d_{11}$. Any path through $d_{01}$
cannot result in a smaller value for cells $d_{1k}$, $2 \le k \le n$, since cells $d_{0,k-1}$ have the
same value. In a similar manner, cells in the second column starting at the third line are dominated by $d_{11}$.
These arguments apply not only to $d_{00}$ but to all $d_{ij}$ in general, proving the following:

\begin{lemma}
\label{lem:match}
A cell $d_{ij}$ is dominated by $d_{i+1,j+1}$ if $a_j = b_i$.
\end{lemma}

Let us now consider what happens when $a_1 \ne b_1$. In this case we can still find dominated cells in the second row
and column, depending on the first matching character position in each. Let us assume that the first
character in $A$ matching $b_1$ is $a_l$, $2 \le l \le n$. All cells $d_{1k}$, $2 \le k \le l-1$ are dominated
by $d_{11}$, for the same reasons that were described earlier. And a similar domination relation exists in the columns.

Before we generalize the dominance relation through a theorem, we will introduce a new scoring scheme to 
take advantage of the indel unavoidability, which will create another optimization criterion, monotonicity
in the rows and columns of certain parts in our graph. For the new scoring scheme and for the rest of the
description of our algorithm, we will divide our matrix into two parts, separated by the main diagonal.
The first part includes the center and the left corner of the matrix, where the second part includes
the right corner of the matrix, together with the main diagonal (which is shared by both parts).
The scoring scheme and the algorithm described further on will be analyzed on the part of the matrix left of the main diagonal,
although all theory works symmetrically on the part right of the main diagonal, by substituting the rows with columns and vice versa.

The new scoring scheme, for the left part of the matrix, is implemented as follows: Every vertical transition (indel)
incurs a cost of $2$, since it strays away from the main diagonal and creates the need of another horizontal
indel to compensate. All horizontal transitions do not carry any cost. The match and substitution costs 
remain $0$ and $1$ respectively. To obtain the edit distance $s$, we add $n-m$ to the value of cell $d_{mn}$. 
The transformation is illustrated through an example in Fig. \ref{fig:score}.

To guarantee the correctness of an algorithm based on that scoring scheme, we will now prove the following lemma:

\begin{lemma}
Under the new scoring scheme, the edit distance of $A$ and $B$ remains unchanged.
\end{lemma}
\begin{proof}
It has already been shown that the edit distance is defined by an optimal path of the fewest possible edit operations carrying a cost,
resulting in the minimum score at $d_{mn}$.
We will prove the following two statements: 
\begin{enumerate}
\item The score obtained from the optimal path remains unchanged and 
\item No other path can lead to a sequence of fewer edit operations and thus a smaller score / edit distance.
\end{enumerate}
To prove the first statement, we note the following: The number of match and substitution transitions in 
the optimal path do not alter the edit distance in the new scoring scheme, since the costs of these operations have not changed. 
With the optimal path starting at diagonal $0$ and ending at diagonal $n-m$, there are $n-m$ indels which
can be omitted from our calculation, since with the new scoring scheme we add these at the end. The only remaining
edit operations to examine are vertical indels left of the main diagonal and horizontal indels right of the main diagonal, which
must be accompanied by compensatory horizontal and vertical indels 
in the respective parts, or the optimal path cannot end up in the main diagonal. Since these indels come in pairs,
with half of them carrying the cost of $2$ and half the cost of $0$ in the new scoring scheme, the final edit
distance remains unchanged.

The second statement follows from the previous arguments, since any path under the new scoring scheme carries the
same cost as before, so a new path with a better score than the previous optimal path score contradicts the optimality
of the latter under the original scoring scheme.
\end{proof}

Since with the new scoring scheme horizontal transitions do not carry a cost, 
the values of cells in every row in the left part of the matrix
are monotonically decreasing. The same holds for the columns in the right part of the matrix, which
leads to the following:

\begin{corollary}
Under the new scoring scheme, the values of cells in rows left of the main diagonal and in columns right of the
main diagonal are monotonically decreasing as the indices of the corresponding cells increase.
\end{corollary}

Let us now consider all cells in a specific row $x$, left of the main diagonal. 
Values on this row are monotonically decreasing
and we only need to keep the information of the first cells from the right where the values are changing
(the leftmost cells of a series of cells with the same value),
since the rest of the cells are dominated (can be reached with $0$ cost from the aforementioned cells).
Now, if we have two consecutive dominant cells $d_{xy}$ and $d_{xz}$ on row $x$, with $y < z$ and
$d_{xy} = d_{xz}+1$, then the value of $d_{xy}$ can be propagated through a transition to row $x+1$ only if a match exists
between $b_x$ and $a_k$, with $y < k \le z$.
In order to be able to locate such matches in constant time, 
we will create lookahead tables for each letter of the alphabet $\Sigma$, which can point to the next
matching character from strings $A$ and $B$. Basically these lookahead tables will be able to answer the question: 
Given a character $c \in \Sigma$ and a position $1 \le k \le n$, what is the smallest index $l \ge k$ such that
$a_l = c$? And the same for string $B$. Such a lookahead table can be easily constructed in time and space 
$O( (n+m)|\Sigma|)$, which for a fixed alphabet of constant size is linear, by traversing both strings in reverse
order, once for each character of the alphabet.

One can easily verify that lemma \ref{lem:match} still holds, based on the same arguments used to prove it,
under the new scoring scheme. In addition, the following corollary holds:

\begin{corollary}
\label{cor:diagonal}
A cell $d_{ij}$ with value $D$ dominates all cells $d_{i-k,j-k}$, $0 \le k \le max(i,j)$ with values $\ge D$.
\end{corollary}
\begin{proof}
It is easy to see, with a simple inductive argument, that a cell $d_{ij}$ dominates all parental cells
on the same diagonal with the same score. Since any cell dominates itself with a higher score (because
every path from that cell will have a higher score equal to the difference of the two scores), the corollary
follows.
\end{proof}

\subsection*{The algorithm}

The algorithm works separately on the two parts of the matrix left and right of the main diagonal.
For the description of the algorithm will consider only the part of the matrix lying left of the main diagonal,
with the assumption that all operations are symmetric on the right part of the matrix. An exception
will occur when we describe the interface of the two parts.

Our edit distance algorithm is score based. On each iteration the edit distance score is incremented by $1$ and
the part of the edit graph that can be reached with the current score is determined. The initial score is $0$, although
we should keep in mind that, since at the end we add $n-m$ to the score -- adjusting for the unavoidable indels
that we get for free on horizontal transitions -- it can be considered as if the score is initialized with the value $n-m$.

\begin{figure*}[htbp]
\centering
\subfloat[Score $0$ iteration]{\epsfig{file=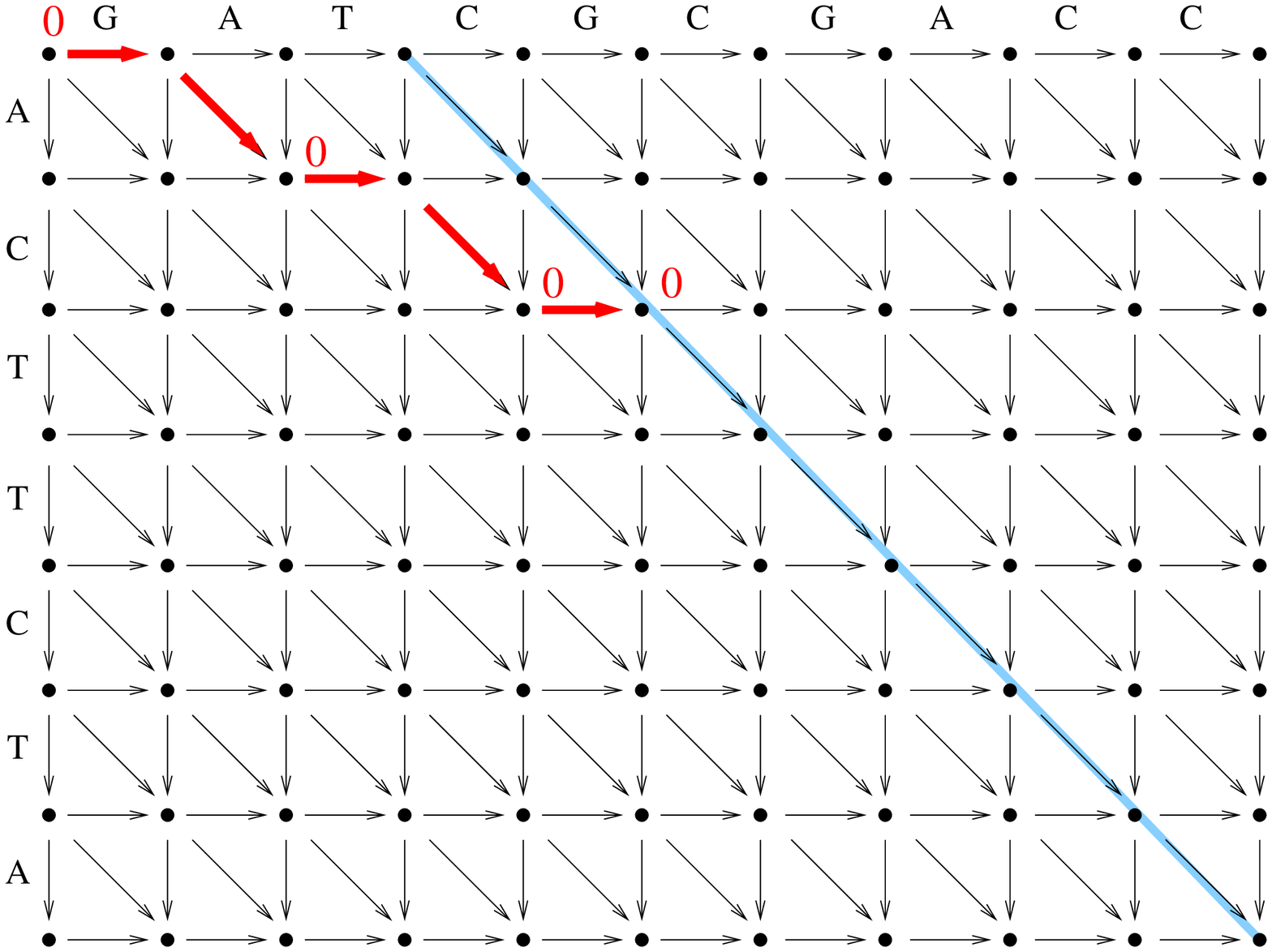, width=0.48\textwidth}}
\subfloat[Score $1$ iteration]{\epsfig{file=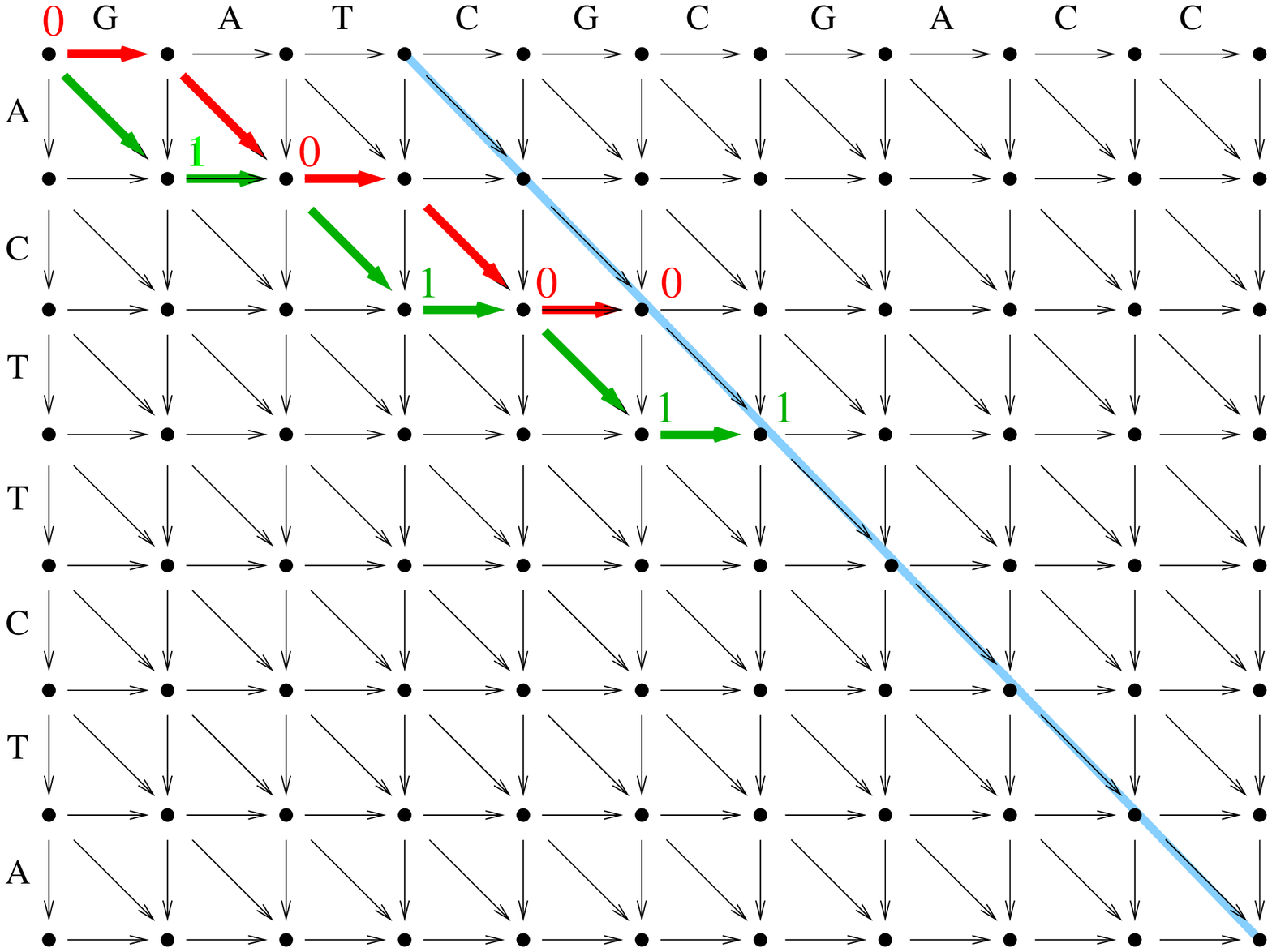, width=0.48\textwidth}}
\subfloat[Score $2$ iteration]{\epsfig{file=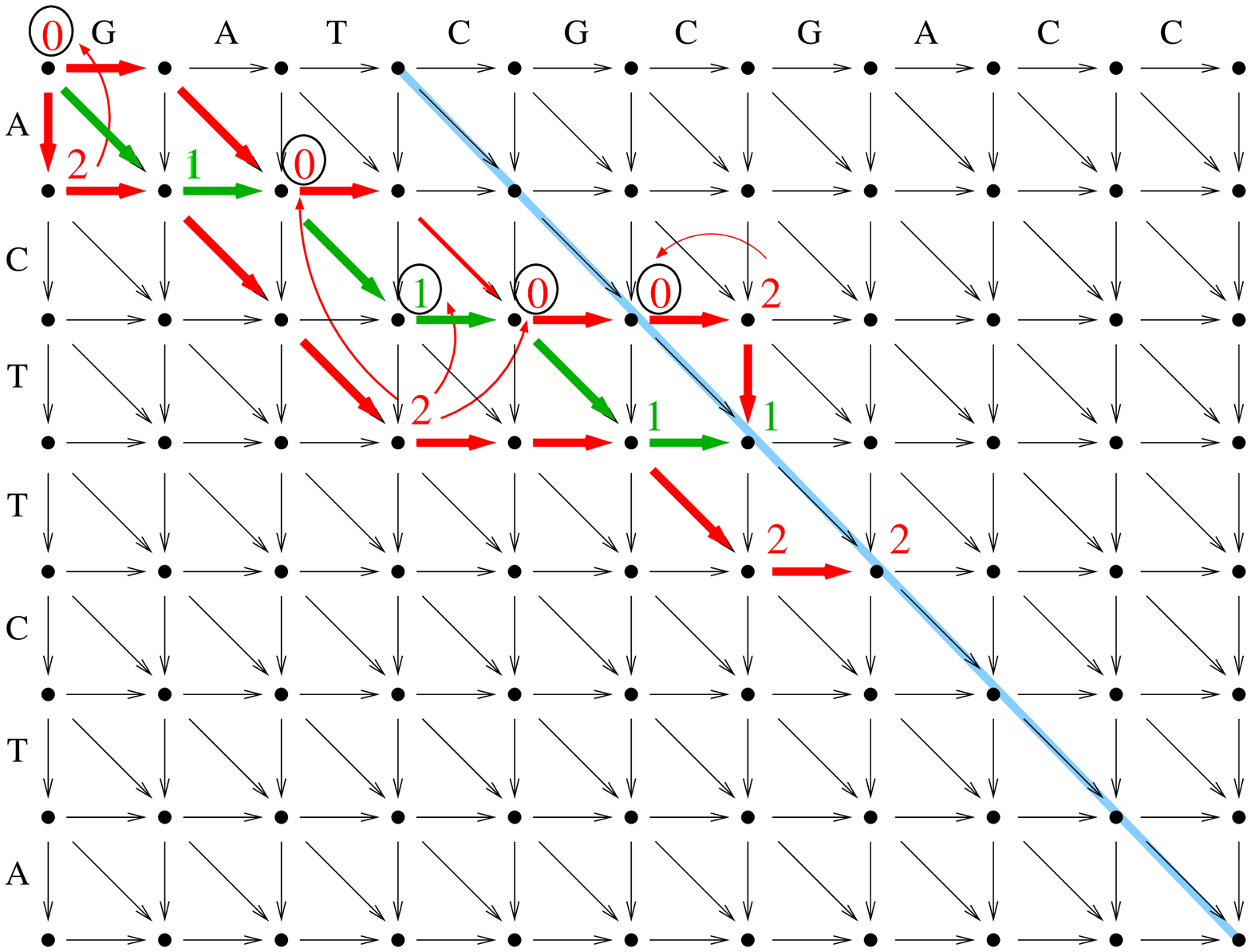, width=0.48\textwidth}}
\subfloat[Score $3$ iteration]{\epsfig{file=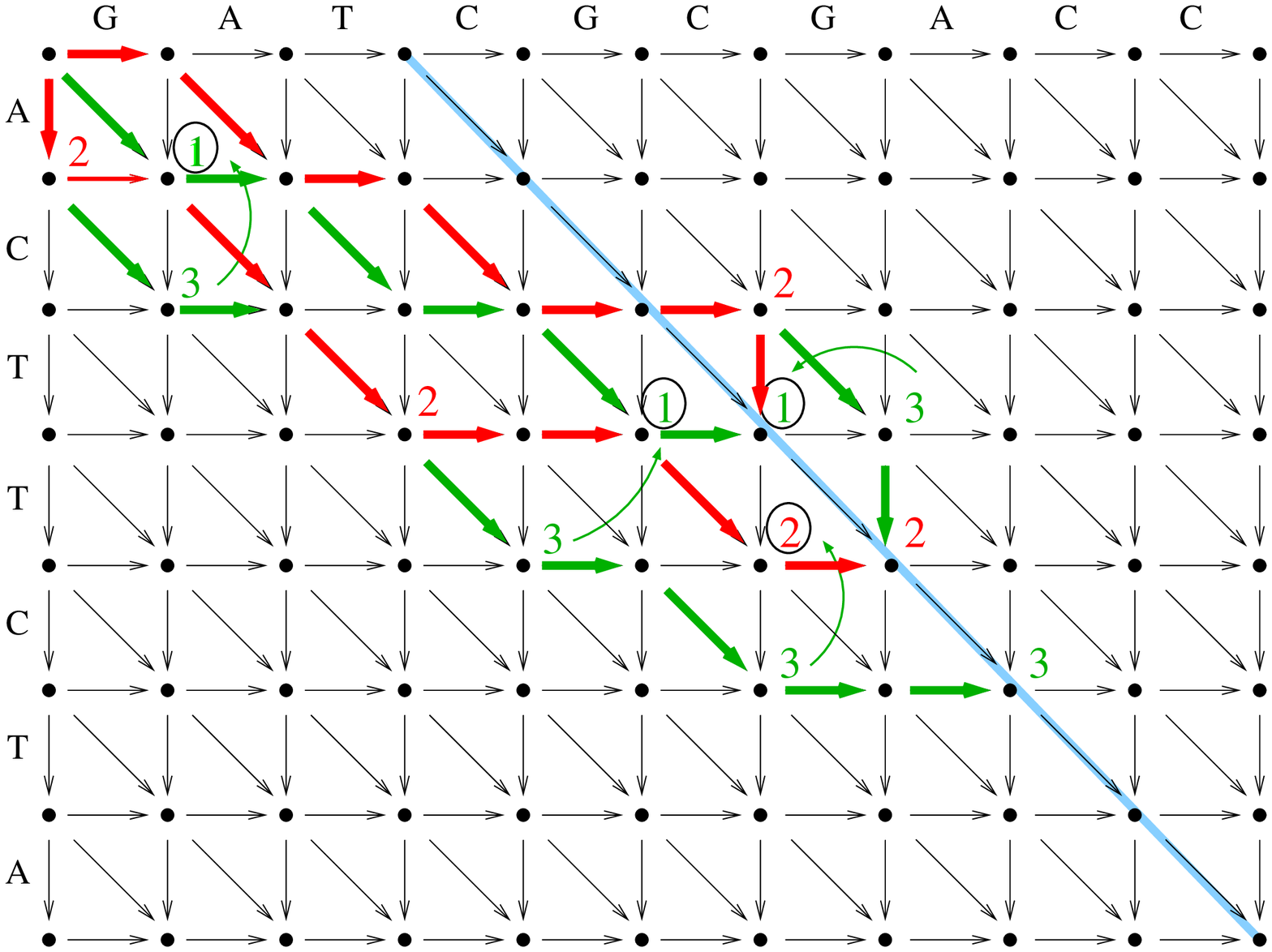, width=0.48\textwidth}}
\subfloat[Score $4$ iteration]{\epsfig{file=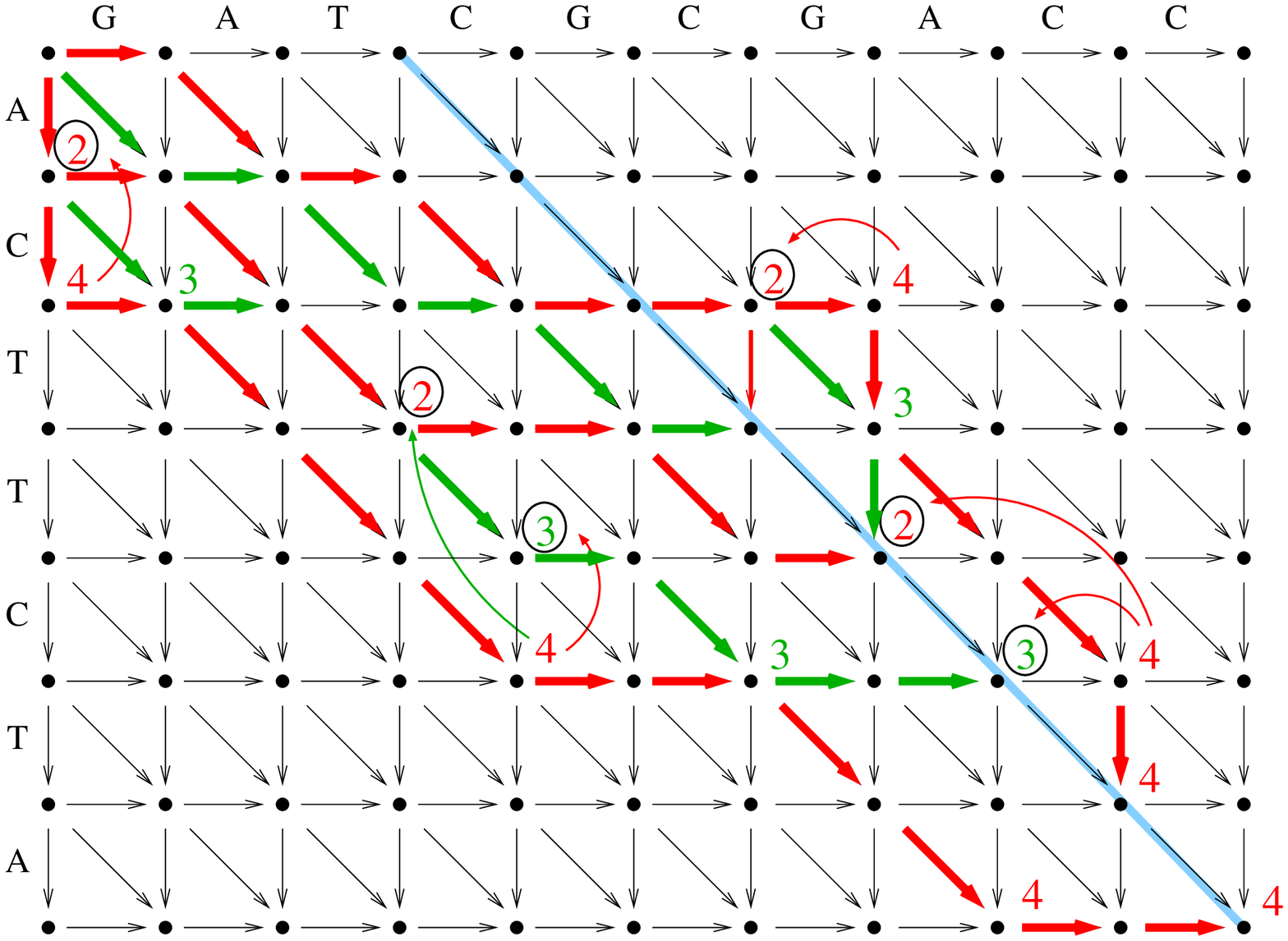, width=0.48\textwidth}}
\caption{Edit distance algorithm iterations. The main diagonal is depicted in blue, iteration transitions are drawn in red and green
alternatively. Cells whose values are presented have been inserted in the list at the end of each iteration, where
cells that their values are circled have been removed from the list, dominated by the cells they connect with arcs.}
\label{fig:algo}
\end{figure*}

During each iteration, we keep the values and positions of the cells we work with in a double linked list,
which will be referred to simply as {\it list}.
To store the position of a cell we actually need only the column index where the cell resides, for
reasons that will be explained later. The initialization phase starts with the determination of the cells
which can be reached with a score of $0$. Since all horizontal and match diagonal transitions (diagonal transitions corresponding to matching characters) have a cost
of $0$, we follow horizontal transitions until we locate a match, then advance to the next line and repeat.
The process ends when we reach the main diagonal. We do not need to keep information on all cells
with $0$ value, the first cell with a value of $0$ on each line suffices, since all further cells
are dominated. These dominant leftmost cells can be located in constant time for each line, by using the lookahead tables.
When we encounter a series of matches on the same diagonal, we only need to keep the value of the last (bottom-right) cell, since
all other cells are dominated. The indices of cells accessed through this process increase monotonically, as
we advance forward through rows, columns and diagonals. The initialization finishes when the main diagonal is reached.
Thus at the end of the initialization step we have a list of cells with $0$ value, each of which resides
on a different row, column and diagonal of the matrix. An example of the initialization phase can be found 
in Fig. \ref{fig:algo}a.

On each subsequent iteration of the algorithm and with each increasing value of the score, 
the linked list is updated with new cells 
that can be reached from members of the list. The algorithm at iteration $D$, with $D$
also being the current score, starts from the top of the list and processes one cell at a time. For each list
cell examined having a value of $D-1$ or $D-2$, as will be proved in lemma \ref{lem:score}, we either follow
a substitution transition, if the cell's value is $D-1$ or a vertical indel transition if the cell's value is
$D-2$. Lets assume we are examining list cell $d_{ij} = D-1$. We know that $d_{i+1,j+1} = D$, since if $d_{i+1,j+1} < D$
it would already be included in the list, unless dominated by another cell in the list, which is impossible
since then $d_{ij}$ would in turn be dominated by $d_{i+1,j+1}$ and would not be in the list during the current
iteration. We now find the largest $k$ for which $b_{i+k} = a_{j+k}$, $k \ge 1$ and insert cell $d_{i+k,j+k}$
in the list. That is the last cell in a series of match transitions, starting at $d_{i+1,j+1}$, if any exist.
Next, we examine the cells following $d_{ij}$ in the list and remove the ones that are dominated by $d_{i+k,j+k}$.
At this step, list cells $d_{op}$ in rows $o < i+k$ and on diagonals $o-p$ such that $i-j < o-p \le n-m$
are removed, all being dominated as proved later in
theorem \ref{thm:row_domination}. Starting now at cell $d_{i+k,j+k}$, we repeat the process performed in the
initialization, with the difference that for each new cell inserted in the list, all subsequent
cells in the list that are dominated by the new member are removed. This process will stop once the next
identified match in the lookahead table falls inside the dominated area. Precisely, if $d_{op}$ is the last
cell with value $D$ that was inserted in the list, the next match from the lookahead tables
resides at diagonal $q$ and the next cell in the list resides at a diagonal $p \le q$ and row $r \ge o$,
then the process of inserting new cells derived from $d_{ij}$ is terminated and we proceed to the next
cell in the list.

Each iteration finishes once we reach the main diagonal. The reader can follow the procedure, through the
five iterations in calculating the edit distance of strings $A = $ 'GATCGCGACC' and $B =$ 'ACTTCTA', 
in Fig. \ref{fig:algo}. 

One special case that was not covered in the above description is the handling of a cell insertion following
a vertical indel transition, when another dominated cell on the same diagonal exists in the list. In this case, 
the only position the dominated cell can occupy is previous to the current cell examined, from which the
transition emanated. This results in the removal of the dominated cell. This special case only requires 
a constant number of operations and does not alter the complexity of the algorithm.

As already mentioned, the part of the matrix right of the main diagonal	is processed in a symmetric way.
At the end of each iteration, the cells of the main diagonal, which belongs to both parts, have to be updated.
These cells reside at the end of the lists for both parts and the update is performed in constant time as well.

We will now proceed to prove the following theorem:

\begin{theorem}
\label{thm:row_domination}
Cell $d_{ij}$ on diagonal $i-j$ with value $D$ dominates all cells $d_{kl}$ 
in the list with $k < i$, $i-j < k-l \le n-m$ and values $< D$, meaning all 
list cells in rows above it and columns with larger indices.
\end{theorem}
\begin{proof}
Since horizontal transitions carry a cost of $0$, all cells in row $i$ and column $l$ with $j < l \le n-m$ 
have a score of at most $D$. All cells $d_{kl}$ in the list, residing in diagonals $k-l$ with 
$i-j < k-l \le n-m$ and in rows $k$ with $k < i$ lead diagonal transitions to cells $d_{k+1,l+1}$ with score at
most $D$, since $a_l \ne b_k$ (or $d_{kl}$ would not belong to the list, dominated by $d_{k+1,l+1}$).
This implies that no diagonal transition from these cells can produce a value smaller than $D$ in
any cell on row $i$ and column $> j$ via a path passing through these cells, 
since values in the paths are monotonically
increasing (because all edit operations have non-negative costs). If we now examine the vertical
transitions emanating the $d_{kl}$ cells under consideration, they also result in paths propagating scores at
least $D$, which again cannot result in a better score on the cells on row $i$ and column $> j$.
All cells on diagonals $< i-j$ do not need to be considered, since they cannot be reached from
the claimed dominated cells of this theorem, unless a path reaches them through a cell in diagonal
$i-j$. But in corollary \ref{cor:diagonal} we showed that cells on this diagonal with scores $\ge D$ are already dominated by $d_{ij}$. Thus all $d_{kl}$ cells are dominated by $d_{ij}$.
\end{proof}

The next corollary follows from the domination theorem \ref{thm:row_domination}:
\begin{corollary}
\label{cor:column}
No two cells in the list reside on the same column. 
\end{corollary}
\begin{proof}
Before a new candidate cell $d_{ij}$ is inserted in the list, any list cell on the same
column will be removed, since it is dominated by the newly inserted cell, based on the
previous theorem.
\end{proof}

Now we have the necessary tools to prove the following lemma:

\begin{lemma}
\label{lem:score}
When iteration $D$ starts, with $1 \le D \le s - (m-n)$, all cells in the 
linked list have either a score of $D-1$ or $D-2$.
\end{lemma}
\begin{proof}
Initially, after the initialization, the list holds cells with value $0$, so the lemma holds. Every
time a cell is inserted in the list it will remain until it is dominated by another cell or
the algorithm terminates. Unless a cell with score $D$ in the list is dominated and 
removed before its transitions are examined,
when the algorithm reaches that cell the diagonal transition emanated from the cell will produce the
next candidate, with score $D+1$, to be inserted in the list. The second time this cell is visited, the
vertical transition from it will be examined. In that case, the next candidate with score $D+2$ will dominate
the current cell, according to the previous theorem. Thus, even if a cell is not dominated by 
another inserted cell, it will be dominated by its siblings.
\end{proof}

A direct consequence of the previous lemma is the following:
\begin{corollary}
\label{cor:score}
At most two cells in the list reside in the same diagonal, and their values differ by $1$. This holds for same row list cells as well.
\end{corollary}

A pseudo-code description of the algorithm is presented below. The description excludes special 
cases requiring substitutions of the currently examined cells of the list and only presents the 
operations of the algorithm in
the part of the matrix left of the main diagonal. The procedure interfacing the left and right linked lists
is omitted as well. The code can be studied in more detail from the available code.

\medskip
\begin{algorithmic}
\STATE Initialize lookahead arrays $X$
\STATE Initialize linked list $L$
\STATE score $D := 0$ 
\STATE line $l := 0$
\STATE column $c := 0$
\WHILE{Not reached main diagonal}
\STATE insert $d_{lx} := $X[$a_l$][$c$] into $L$
\STATE $c := x$
\STATE $l++$
\ENDWHILE
\medskip
\WHILE{Not reached cell $d_{mn}$}
\STATE $D++$
\STATE Current Cell $d_{ij} := L\rightarrow$head
\REPEAT
\IF{$d_{ij} = D-1$}
\STATE $d_{ij} := ${\it process\_next\_candidate}($d_{i+1,j+1}$)
\ELSE
\STATE $d_{ij} := $ {\it process\_next\_candidate}($d_{i+1,j}$)
\ENDIF
\UNTIL{$d_{ij} = L\rightarrow$head}
\ENDWHILE
\medskip
\STATE {\bf Function} process\_left\_candidate($d_{kl}$)
\WHILE{$a_{l} = b_{k}$}
\STATE $k++$
\STATE $l++$
\ENDWHILE
\STATE Insert $d_{kl}$ in list $L$
\STATE Remove dominated $d_{ij}\rightarrow$next by $d_{kl}$ from $L$
\WHILE{not reached diagonal of $d_{ij}\rightarrow$next}
\STATE {\it process\_left\_candidate}(X[$a_{k+1}$][$l+1$])
\ENDWHILE
\RETURN $d_{ij}\rightarrow$next
\end{algorithmic}
\medskip

\subsection*{Algorithm complexity}

The algorithm described in the previous section is score based and as such the main loop executes an equal
number of times with the value recorded at cell $d_{mn}$ of the edit graph. Since we add the value $n-m$ to
that score in order to obtain the edit distance of strings $A$ and $B$, the total number of iterations
is equal to $s-|n-m|$.

At all times during the execution of the algorithm the linked list contains at most $m$ cells, which is a 
direct consequence of corollary \ref{cor:column}. Also, due to corollary \ref{cor:score}, there can be
at most $2s$ cells in the list at any given time, since the maximum number of diagonals on which the algorithm
processes cells is $s$, consisting of the center of the matrix and diagonal bands of size $(s-|n-m|)/2$ from
each side of the center, accessed while the algorithm iterates. Basically, for every two iterations of the algorithm,
one further diagonal from each side of the center of the matrix is accessed.

All cells in the list are accessed in order and without backtracking during each iteration. Each cell undergoes
through a constant number of structural accesses, once when it is inserted in the list, once when it is removed and two
times when the diagonal and vertical transitions from this cell are examined, if there is a chance before
it is dominated. During each iteration there are other cells accessed, the candidates for insertion
in the list. While processing these cells we are advancing both the indices of columns and rows
without backtracking,
which proves, as with list cells, that there are at most $m$ or $s$ candidate cells examined during each iteration.

A candidate cell may be accessed several times while compared to a list cell, in
order to determine a dominance relation. A list cell can also be accessed several times during
the same process, to check whether it is dominated. 
The amortized cost for each cell though is constant. Every time a candidate cell is re-examined, a cell
from the list has been removed. And every time a list cell is re-examined, in the previous step it 
was not dominated by a candidate cell, the latter then having being inserted in the list
and not being examined again on that iteration. Since each time we
advance through either a candidate or a list cell, and since both sets have
$O(min(m,s))$ cells (under the assumption that $m \le n$), the total number of constant time operations 
during an iteration is $O(min(m,n,s))$.

This analysis demonstrates that the total running time of our algorithm is $O( (s-|n-m|)\cdot min(m,n,s) + m + n)$,
where the last linear $m+n$ component represents the time necessary to initialize the lookahead tables. 
It can be easily verified using simple algebra that $s-|m-n| \le min(m,n)$, which provides
another less tight upper bound of the worst case time behavior of the algorithm,
$O(min^2(m,n,s)+m+n)$.
We can therefore observe
that the quadratic factor in the time complexity is independent of the longest string being compared.
The space usage of this algorithm is $O(m+n)$, dominated by the size of the lookahead tables kept in memory. 
This completes the proof of the next theorem:

\begin{figure*}[htbp]
\centering
\subfloat[Strings on $4$-letter alphabet]{\epsfig{file=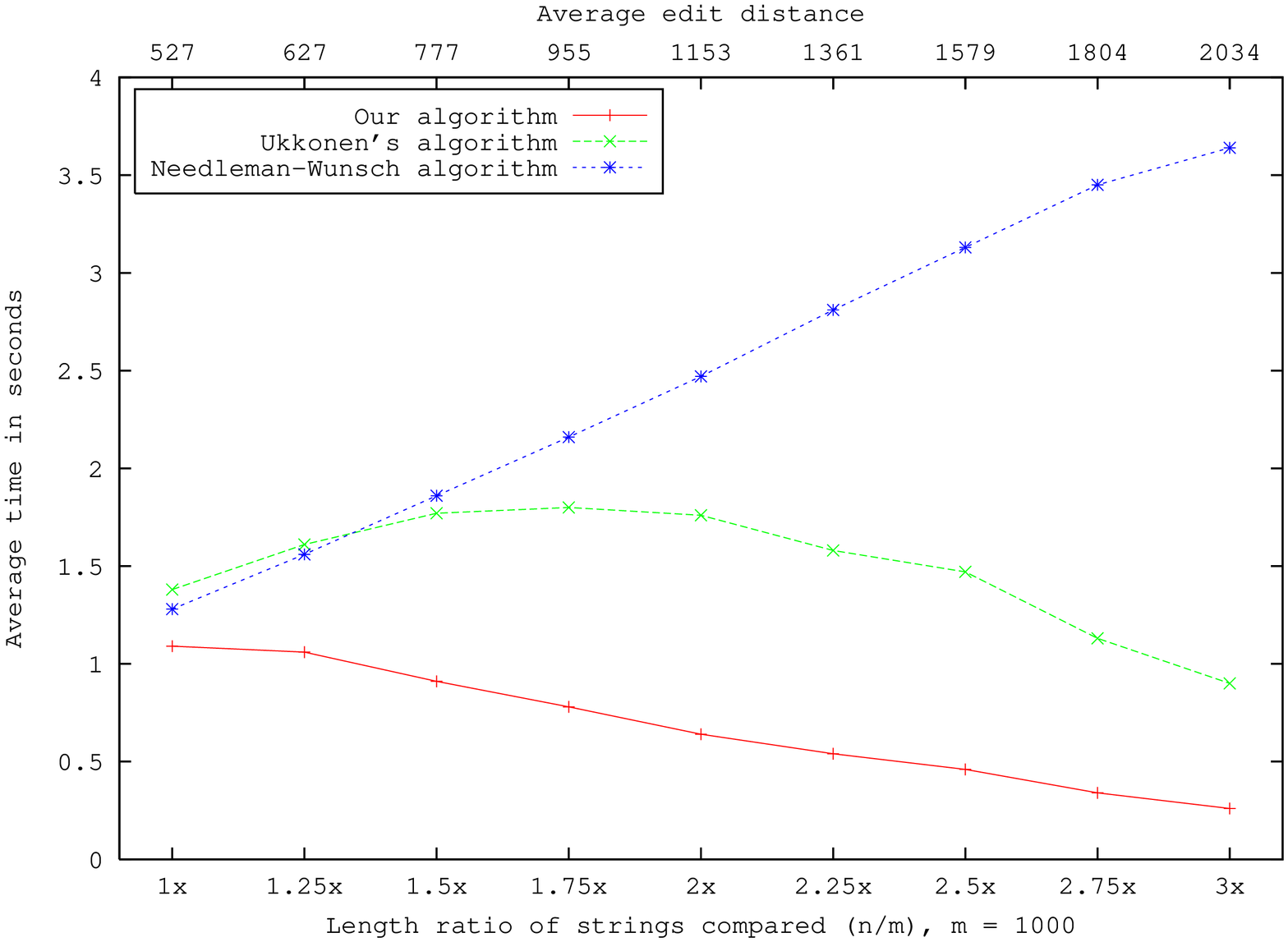, angle=-90, width=0.48\textwidth}
}
\subfloat[Strings on $20$-letter alphabet]{\epsfig{file=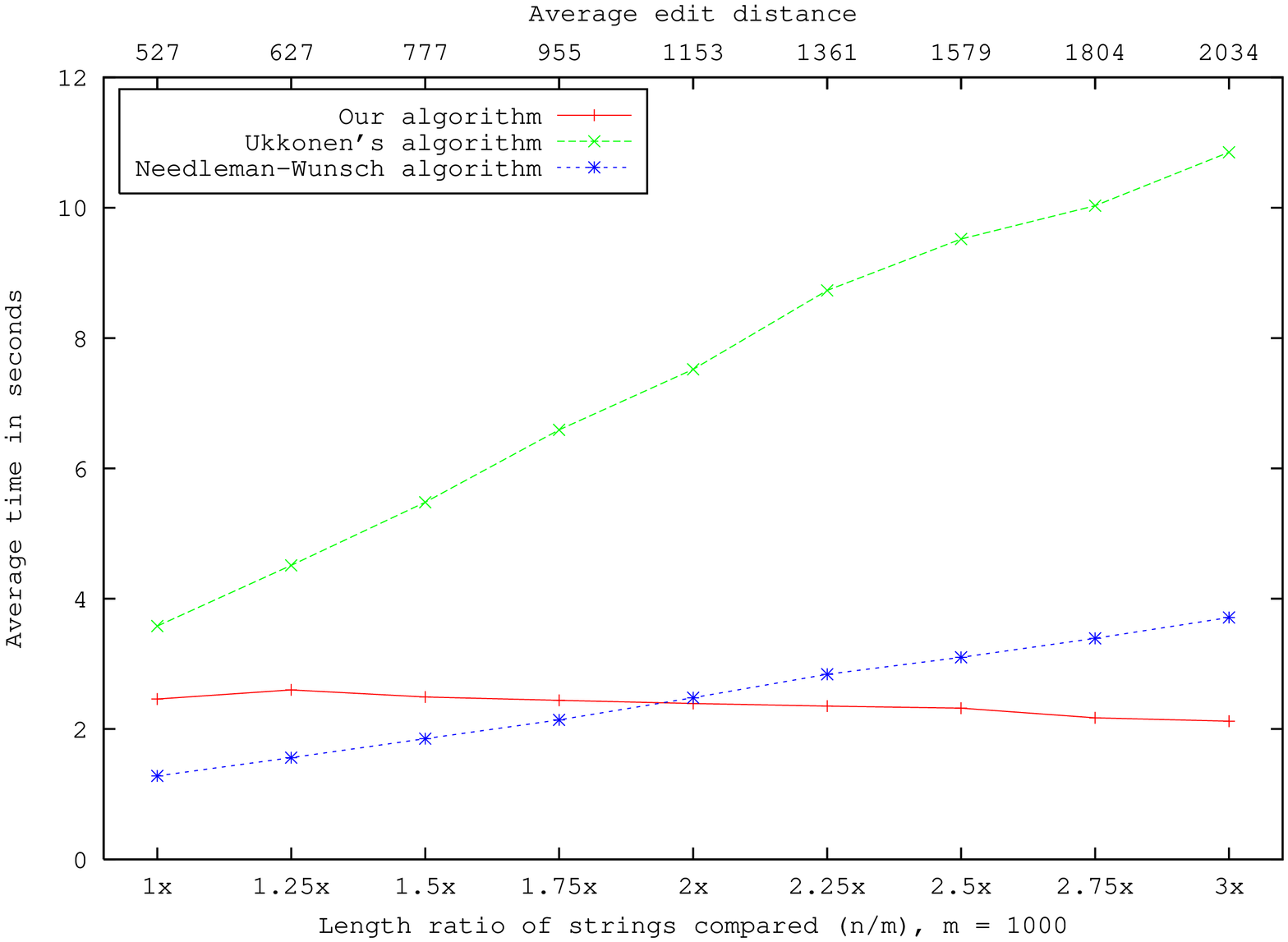, angle=-90, width=0.48\textwidth}
}
\caption{Edit distance calculations on random strings with different length ratios, comparing the performance of our, Ukkonen's and the basic algorithms}
\label{fig:random}
\end{figure*}

\begin{theorem}
The edit distance $s$ of two strings $A$ and $B$ with lengths $n$ and $m$ respectively can be computed
in time $O( (s-|n-m|)\cdot min(m,n,s) + m + n)$ and in space $O(m+n)$.
\end{theorem}

\section*{Results}

We have implemented our new algorithm to test its performance in practice.
For comparison purposes, we implemented the basic $O(mn)$ algorithm, also known as Needleman-Wunsch \cite{nee70},
as well as the Ukkonen $O(s\cdot min(m,n))$ algorithm \cite{ukk85}. All algorithms were implemented in 
perl, using the same input/output procedures and no optimizations. Benchmarking was performed with the 
{\it benchmark} perl module for the experiments averaging a large number of random runs, and the {\it time} unix command
for individual experiments, the same method always used across algorithms. 
All tests were performed on an 8GB RAM 2.93GHz Intel processor IBM compatible
desktop machine, running ubuntu linux. In all test cases the data completely fit in the main memory.

Since perl does not support pointer structures efficiently, we implemented the double linked list with
arrays, using the fact that no two cells in the list can reside on the same column. 
This way we access list cells using their column index. As such, the list
occupies more space than the minimum possible, where the implementation may have been more efficient 
in another programming language supporting these structures.

Ukkonen's algorithm implementation was based on the outline found in \cite{ukk85} and the more detailed
description found in \cite{pow01}. The version used is particularly simple by making use of recursion,
but has larger than linear space demands, specifically $O(s\cdot max(m,n))$.
The basic algorithm was implemented using linear space and row-by-row iterations.

The first two experiments were run on random sequences over alphabets of $4$ and $20$ characters respectively,
similar to random DNA/RNA and amino acid sequences.
The length of the first sequence from the two compared was set at $1000$ characters, where the length of the second sequence varied between
$1000$ and $3000$ characters. We examined a total of nine length ratios $n/m$ values between $1$ and $3$ ($1 \le n/m \le 3$). 
For each length ratio, $100$ different comparisons were run, with the execution time and 
edit distance values averaged among these. The results
are depicted in Fig. \ref{fig:random}.

\begin{table*}
\scriptsize
\centering
\begin{tabular}{|c|c|c|c|ccc|c|}
\hline
\multirow{3}{*}{Sequence A}&\multirow{2}{*}{Sequence B}&\multirow{2}{*}{Alphabet}&\multirow{2}{*}{(Average)}&Our&Ukkonen's&Basic&(Average)\\
&&\multirow{2}{*}{size}&\multirow{2}{*}{length}&algorithm&algorithm&algorithm&edit\\
&&&&(sec)&(sec)&(sec)&distance\\
\hline
\hline
Random 16S&Random 16S&\multirow{2}{*}{4}&\multirow{2}{*}{1350}&\multirow{2}{*}{0.679}&\multirow{2}{*}{0.811}&\multirow{2}{*}{2.554}&\multirow{2}{*}{421.3}\\
rRNA sequence&rRNA sequence&&&&&&\\
\hline
Hyphomonas 16S&Hyphomonas 16S&\multirow{2}{*}{4}&\multirow{2}{*}{1330}&\multirow{2}{*}{0.25}&\multirow{2}{*}{0.18}&\multirow{2}{*}{2.14}&\multirow{2}{*}{46}\\
rRNA (AF082798)&rRNA (AF082795)&&&&&&\\
\hline
Alphaproteobacteria 16S&Betaproteobacteria 16S&\multirow{2}{*}{4}&\multirow{2}{*}{1320}&\multirow{2}{*}{0.42}&\multirow{2}{*}{0.46}&\multirow{2}{*}{2.07}&\multirow{2}{*}{318}\\
rRNA (AJ238567)&rRNA (AJ239278)&&&&&&\\
\hline
Cucumber necrosis&Lisianthus necrosis&\multirow{2}{*}{4}&\multirow{2}{*}{4790}&\multirow{2}{*}{6.70}&\multirow{2}{*}{6.32}&\multirow{2}{*}{28.27}&\multirow{2}{*}{1154}\\
virus genome&virus genome&&&&&&\\
\hline
Human poliovirus 1&Human Rhinovirus A&\multirow{2}{*}{20}&\multirow{2}{*}{870}&\multirow{2}{*}{1.02}&\multirow{2}{*}{1.05}&\multirow{2}{*}{0.88}&\multirow{2}{*}{472}\\
virion protein&virion protein&&&&&&\\
\hline
\end{tabular}
\caption{Algorithm performance comparing biologically related sequences of similar length}
\label{tab:compare}
\end{table*}

In these simulations it is worth noticing significant performance improvement of the new algorithm with increasing length ratio
of the random strings, although the total length of the strings is increasing. This is not surprising, since the
number of iterations $s-|m-n|$ is decreasing, caused by a slower increase in edit distance than difference
between the lengths of the two strings.

Ukkonen's algorithm is under-performing when comparing random strings over a large alphabet, because
of the large expected edit distance value in these cases. This algorithm is designed for comparing similar
strings, which is the case most often encountered in practice. In contrast, the basic algorithm,
owing to its simplicity, is performing consistently and surpassing the other algorithms when
the edit distance is large compared to string length, unless when the $s-|n-m|$ value becomes small
enough, where our algorithm takes the lead.

Next, we designed computational experiments performing comparisons most often encountered in practice,
drawn from the computational biology domain. In all examples the sequence pairs examined have
comparable lengths, not differing more than $5\%$. The results are presented in Table \ref{tab:compare}.
The first simulation involved $1000$ random sequence pair comparisons from
a pool of approximately 6800 vetted 16S ribosomal RNA sequences, provided by the Ribosomal Database Project (RDP) at
http://rdp.cme.msu.edu. These sequences average about $1350$ characters in size, drawn from an alphabet
of size $4$. A random pair of 16S rRNA sequences from the same genus and another from the same class
but different order are compared in the next two lines, followed by a comparison of two viral genomes
and two virion proteins.

As these results demonstrate, the performance of our algorithm compares favorably to Ukkonen's asymptotically
slower but with lower constants algorithm, while the basic algorithm is outperformed in almost every
case, except when matches are sparse. Performance comes with some cost though and it is interesting 
to note that the size of the program implementations
of the three algorithms, the basic, Ukkonen's and ours, is $80$, $160$ and $700$
lines of code respectively.

The perl implementations of all three algorithms used in this paper for
performance comparisons can be downloaded at http://www.cs.miami.edu/\~{ }dimitris/edit\_distance.

\section*{Conclusions}

In this paper we have provided the design, analysis and implementation of a new algorithm for calculating
the edit distance of two strings. This algorithm is shown to have improved asymptotic time behavior, while
it is also demonstrated to perform very well in practice, especially when the lengths of the strings
compared differ significantly. The performance of our algorithm in this case, which is encountered 
less often in instances of the edit distance problem, could find application in the related Longest Common
Subsequence (LCS) and other similar problems solved with dynamic programming techniques.

Future directions for this algorithm include the investigation of further practical applications of the
techniques described in other similar problems, as well as generalizing the results for arbitrary 
weights of the edit operations and/or covering additional edit operations such as swaps.


\section*{Acknowledgements}

We would like to thank Gonzallo Navarro, Amihood Amir and Gad M. Landau for information
provided regarding the complexity of the edit distance problem.


{\ifthenelse{\boolean{publ}}{\footnotesize}{\small}
 \bibliographystyle{bmc_article}  
  \bibliography{bibliography} }     


\ifthenelse{\boolean{publ}}{\end{multicols}}{}

\end{multicols}
\end{singlespace}
\end{document}